\documentclass[aps,pra,showpacs,showkeys,onecolumn,notitlepage,groupedaddress]{revtex4-2}
\usepackage[english]{babel}
\usepackage{comment}
\setcitestyle{numbers,square}
\usepackage{array}
\usepackage{amsmath}
\usepackage{units}
\usepackage{graphicx}
\usepackage{braket}
\usepackage{xfrac}
\usepackage{enumitem}
\usepackage[normalem]{ulem}
\usepackage{booktabs}
\usepackage{tabularx}
\usepackage{tikz}
\usepackage{multirow, bigstrut}
\usepackage{hhline}
\usepackage{color,soul}

\usepackage{times}

\usetikzlibrary{shapes.geometric, arrows, matrix, positioning}
\usetikzlibrary{fit}
\tikzset{%
highlight/.style={rectangle,rounded corners,draw,
fill opacity=0.5,thick,inner sep=0pt}
}

\usepackage{hyperref}
\hypersetup{
colorlinks=true,
linkcolor=cyan,
filecolor=magenta, 
urlcolor=blue,
}

\begin{document} 

\title{Building Bridges in Quantum Information Science Education: Expert Insights to Guide Framework Development for Interdisciplinary Teaching and Evolution of Common Language}


\author{Liam Doyle, Fargol Seifollahi and Chandralekha Singh}

\affiliation{
Department of Physics and Astronomy, University of Pittsburgh, Pittsburgh, PA, 15260 USA}

\begin{abstract} The rapid growth of quantum information science and technology (QIST) presents unique educational challenges as it brings together students and researchers from physics, computer science, engineering, chemistry and mathematics. This work presents findings from in-depth interviews with leading quantum researchers who are also educators, whose perspectives provide guidance for developing a framework for interdisciplinary QIST teaching and builds on our earlier paper that focused on QIST courses and curricula educators had taught. We discuss quantum educators' reflections on three critical aspects of QIST education: (1) the development of a common interdisciplinary language, (2) determining appropriate levels of abstraction and physical detail for diverse student populations from various disciplines, and (3) why students should pursue courses, degrees, and careers in this rapidly evolving field. Our analysis reveals that the continuous emergence of linguistic evolutions such as “qubits” and  ``measurement bases", rather than a focus on measurement of physical observables and their corresponding Hermitian operators, has begun to create a unifying framework that transcends disciplinary boundaries. Nevertheless, educators face ongoing challenges in balancing the level of abstractness with physical details as well as mathematical rigor with conceptual accessibility, particularly when teaching foundational QIST courses to an interdisciplinary group of students, who have not taken linear algebra or differential equations and have no prior college physics background. The experts emphasize that successful QIST education for an interdisciplinary student body not only requires a shift from traditional quantum mechanics pedagogy for physics majors, 
but careful consideration of students' diverse prior conceptual and mathematical foundations overall. They unanimously encourage students to pursue QIST related courses, degrees, and careers, highlighting the unique historical opportunity to participate in creating transformative quantum technologies while developing transferable skills for an evolving technological landscape. These findings provide valuable guidance for developing a framework for interdisciplinary QIST teaching especially useful for early foundational courses to prepare students with diverse disciplinary backgrounds for the quantum information revolution.

\end{abstract}


\maketitle


\section{Introduction and Framework}

The 21st century has witnessed a remarkable transformation in how we understand and harness quantum phenomena, giving rise to quantum information science and technology (QIST), an interdisciplinary field that promises revolutionary advances in computing, networking, and sensing by harnessing quantum superposition and entanglement \cite{Preskill,european,raymer2019,flagship,alexeev2021quantum,divincenzo,lloyd,daley2022,altmansimulation,logicalqubit,simulationfeynman,advantage}. This “second quantum revolution” differs fundamentally from the first quantum revolution of the 20th century, which gave us technologies such as lasers and transistors. While the first quantum revolution leveraged quantum mechanics to understand and control properties of matter and light, the second quantum revolution focuses on unprecedented control and manipulation of individual electrons and atoms to process information in fundamentally new ways for diverse applications \cite{aaronson2013, levymrs, qkd2017,trapped,semiconductorqubit,2020superconducting,circuitqed,neutralatom,kimble,awschalom,photon,entanglementphysics}.

The interdisciplinary nature of QIST presents both unprecedented opportunities and significant challenges for education \cite{fox2020cu, singhasfaw2021pt, asfaw2022ieee,muller2023prperworkforce, qtmerzeletal,bitzenbauer3,weissmanphysics,nvcenter1,nvcenter2,jeremytpt,jeremyajp,Benlarmorajp2025}. Unlike traditional quantum mechanics courses designed primarily for physics majors, QIST draws students from computer science, engineering, chemistry, mathematics, and other fields, each bringing different mathematical backgrounds, relevant conceptual knowledge, and professional goals. This diversity enriches the field, but also complicates the educational landscape \cite{kohnle2013, rodriguez2020designing, goorney2024framework, bungum2022quantum, michelini2023research, singh2015review, bondani,donhauser2024empirical}, raising fundamental questions about curriculum design \cite{meyer2022cu,fargol}, pedagogical approaches, and the very language we use to communicate quantum concepts.

The educational challenges in QIST \cite{beckphoton,singh2007comp,marshman2016ejpphoton,Kohnle_2017,devore2020qkd,maries2020mzidouble,kiko,singh2022tpt, qtmerzel, hennig2024new, qtbrang, qthellstern, qtgoorney,qtmeyercu,qtsun,hubloch,hu2023prper,ghimire2025epj,lopez2020encrypt, chhabra2023undergraduate, michelini2022, schalkers2024explaining,marshman2015,justicemathphysics}  
only partly emanate from figuring out the relevant quantum mechanics concepts for a given QIST course and how to teach them to non-physicists from a variety of disciplines effectively. The field itself is evolving with a new conceptual vocabulary that strives to bridge disciplinary boundaries and can play an important role in developing a framework for QIST education, especially valuable for foundational courses. Traditional quantum mechanics courses, with their emphasis on wave mechanics, differential equations, and physical observables and their corresponding Hermitian operators, may not provide the most effective and efficient pathway for students whose primary interest lies, e.g., in quantum algorithms \cite{shor1994,grover,vaziranib,simon}, quantum error correction \cite{shorcode,errorcorrection1,errorcorrection2,surfacecode,kitaev}, or quantum cryptography and related fields \cite{bennett1984,ekert1991quantum,kwiatqkd,RevModPhysqkd,qkdpanchina2017,galvezqkd,Kohnleqkd,woottersqkd,kellyqkd,netoqkdoutreach,nocloning}. Moreover, the rapid pace of technological development in quantum computing, quantum communication, and quantum sensing means that educational programs must prepare students not just for current technologies but for a rapidly evolving landscape.

A few years ago, educators at a US conference brainstormed strategies for developing QIST courses and curricula that could serve undergraduates from different disciplinary backgrounds while maintaining rigorous standards \cite{asfaw2022ieee}. These discussions highlighted the need for a common educational framework that could unite physicists, computer scientists, chemists, engineers, mathematicians and others around shared quantum concepts and applications \cite{nielsen2010quantum,mermin,QISresource,galvez2014resource,raymerbook,dancing,quantumoptics}. Recent efforts to address these challenges have included the development of new interdisciplinary courses and degree programs \cite{meyer2022cu,qtmeyercu,fargol}.

The rapid advancement of QIST as an interdisciplinary field has also driven linguistic innovations. The word ``quantum" as a subject matter is being used as a noun (students learn quantum) instead of adjective as in ``quantum mechanics" or ``quantum physics". Terms like “quantumly” (replacing the more traditional “quantum mechanically”), “measuring qubits” (rather than measuring physical observables), and “measurement basis” (with a focus on the basis in which measurement is performed as opposed to the physical observable measured) have emerged as part of a simplified vocabulary that prioritizes conceptual clarity over traditional physics terminology appropriate for QIST stakeholders from diverse interdisciplinary backgrounds. In particular, this linguistic evolution reflects a deeper conceptual shift. For example, where traditional quantum mechanics courses might emphasize measuring the eigenvalues of Hermitian operators corresponding to observables, QIST education focuses on the eigenvectors that define measurement bases without reference to underlying physical observables, reflecting the emphasis on the information-theoretic perspective that dominates QIST. 

This paper builds on research focusing on courses and curricula that quantum educators had developed that was discussed in a prior publication \cite{fargol}. Here we focus on educators' perspectives useful for the development of a framework for QIST education at this early time in the second quantum revolution, which can be particularly valuable for foundational QIST courses. Five of the educators who contributed to the discussion of courses and curricula in the prior publication \cite{fargol} also had valuable reflections on questions related to developing a framework for quantum education discussed here. This paper presents findings from these extensive interviews with leading quantum researchers who are actively engaged in quantum education, and addresses three critical issues facing the field as our main research questions:

\begin{itemize}
    \item [RQ1.]\textbf{Evolution of a common language:} How can we develop a common language for teaching interdisciplinary QIST courses that serves students from diverse backgrounds equally well?
    \item [RQ2.] \textbf{Determining appropriate level of detail:} What level of physical detail vs. abstraction is appropriate for an interdisciplinary QIST course for students at a certain level?
    
    \item [RQ3.] \textbf{Experts' messages to students interested in QIST:} What guidance should we provide to students considering courses, degrees, and careers in this rapidly evolving field?
\end{itemize}

Thus, unlike the earlier paper that focuses on specific courses and curricula developed by educators \cite{fargol}, this research work focuses on quantum educators' thoughts on higher level issues in an effort to build a framework for interdisciplinary QIST education, including how to develop a common language and find the right balance between physical details and abstraction. These issues bring out the tension within the interdisciplinary QIST education from two opposing view points. 
This tension is particularly evident currently in the early stages of the second quantum revolution, when many qubit architectures are competing to build fault tolerant quantum computers. In particular, in classical computers, the quantum mechanical nature of transistors is so well understood that an individual can have an entire career in the semiconductor industry without knowing any quantum concepts. That is not the case in QIST, in which the details of qubits are at the forefront and no qubit architecture has succeeded in building fault tolerant quantum technologies. Therefore, learning some physical basis underlying the concepts (e.g., the Controlled NOT gate) is likely to be important for an interdisciplinary student body to understand the physical details in this noisy intermediate state quantum (NISQ) era \cite{Preskill}. However, without a framework for QIST education at this time, how much detail of the physical basis would be appropriate and should be included in a QIST course at a particular level (especially foundational courses) and what should be the learning objectives and goals remain to be determined.

This research uses Vygotsky's framework stressing the Zone of Proximal Development (ZPD) \cite{vygotsky1978mind,WassSharpeningZPD}. The ZPD is defined by what students can do on their own vs. what they can do with the help of educators who are familiar with their prior knowledge and skills. Educators who are familiar with students' prior knowledge and skills can remain within their students' ZPD and scaffold their learning by providing them appropriate support that builds on their prior knowledge. These considerations are particularly important in interdisciplinary QIST education since students from diverse backgrounds and disciplines are likely to be in the same course at a given level. Therefore, it is urgently necessary to develop a framework for QIST education which establishes a common language that is not too heavy in physical basis, while balancing the level of abstraction with some physical underpinnings of foundational concepts at this early stage of the growth in QIST. 

Through think-aloud interviews with several quantum educators, who are leading quantum researchers, we explore experienced practitioners' perspective on these issues to develop a framework for QIST education. Our findings reveal both convergent themes such as the unifying power 
of already-developed common terms such as qubits, and the ongoing tensions between abstraction and understanding of the underlying physical basis of QIST concepts in interdisciplinary  courses, e.g., on foundations of QIST. The insights shared by these educators provide valuable guidance for the development of a framework for interdisciplinary QIST education at this early time in the second quantum revolution, while also benefiting the broader QIST education community in navigating its challenges. 

\section{Methodology}

This research is part of a broader investigation, in which we conducted individual interviews lasting 1 hour-1.5 hours with 13 quantum educators, who are leading researchers in QIST. We asked them many questions related to QIST focusing on different facets. For example, we previously reported their views on misinformation in QIST \cite{kashyap2025strategies}, how to diversify QIST \cite{ghimire2025reflections}, QIST courses and curricula they had developed \cite{fargol}, and current state and future prospects of quantum technologies \cite{liam}. This paper is synergistic and complements these earlier findings and focuses on investigation of expert perspectives for developing a framework for QIST education that can be particularly valuable for early foundational QIST courses. As mentioned earlier in the introduction, three main research questions in this work are focused on 1) evolution of a common language, 2) determining appropriate level of physical detail, and 3) experts' messages to students interested in QIST. Using semi-structured interviews, the following questions were asked to educators to capture their insights in detail about issues in QIST education:

\begin{itemize}
    \item [1.] QIST is a relatively new field. How do we come up with common language to teach interdisciplinary courses involving both physicists and non-physicists so that all students benefit equally, and how do we educate/communicate effectively with diverse stakeholders in this field?

    \item [2.]How do we determine the level of physical detail appropriate in an interdisciplinary QIST course? [\textit{Educators were further provided with the following example to clarify the intent of the question and avoid ambiguity of responses}] For example, students in one such undergraduate interdisciplinary course on foundations of quantum computing and quantum information were asked ``Can a multiqubit state be an entangled state without any prior interactions between the qubits (either directly between two qubits or via other qubits)? Explain \cite{Peterhucompute}." Many students struggled on this question after lecture-based instruction with a number of students stating that entangled qubits need not have prior interactions (direct or indirect), with some saying that simply passing them through an appropriate gate would entangle them. Do you think this level of understanding is appropriate in such an interdisciplinary undergraduate course focusing on foundations of QIST?

    \item [3.] What is your message for students who are interested in interdisciplinary certificates and degrees and wondering whether they should join the workforce to be part of the second quantum revolution?

\end{itemize}

\begin{figure}
    \centering
    \includegraphics[width=0.45\linewidth]{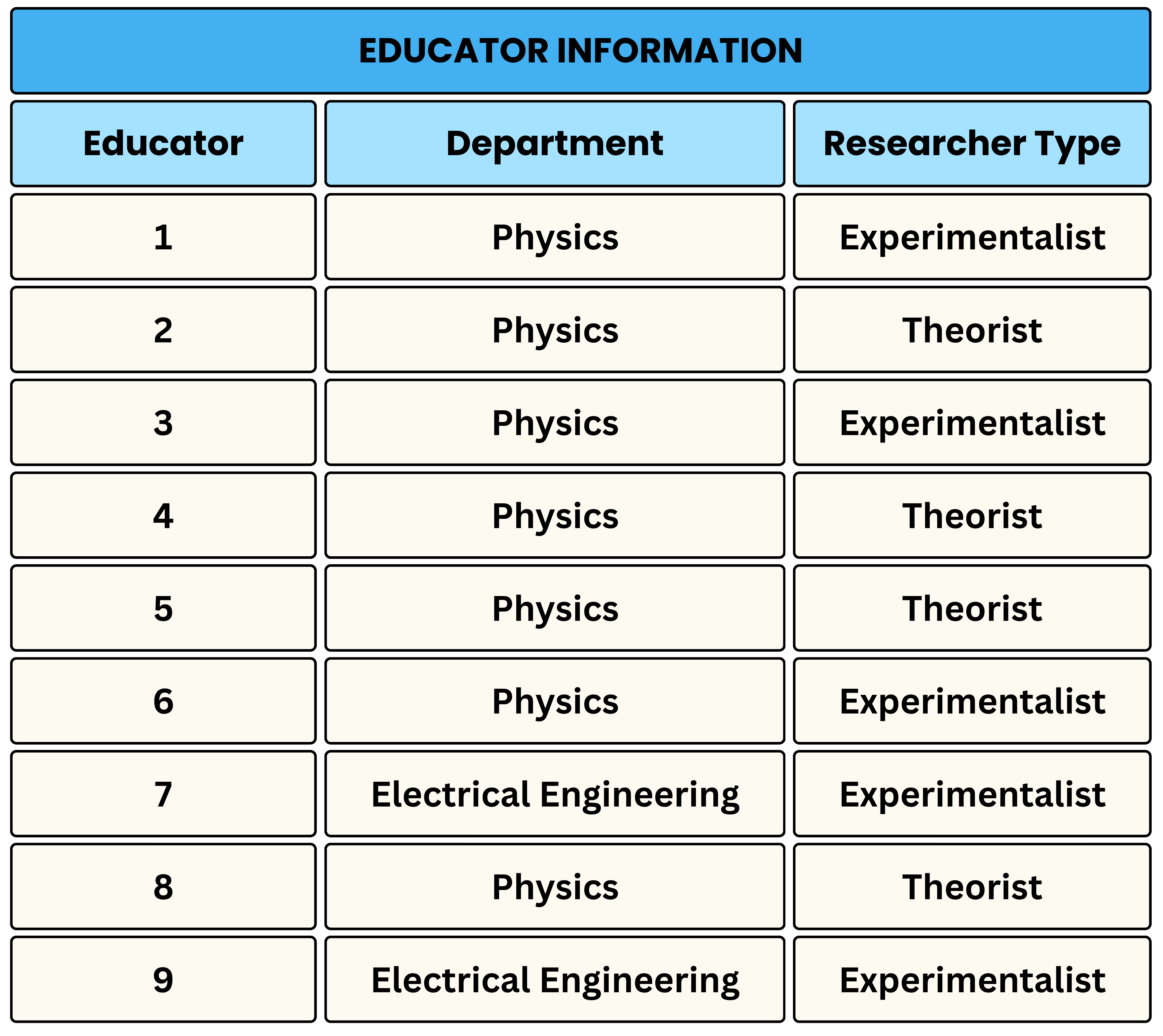}
    \caption{This figure provides information on the nine out of thirteen educators present in the final study sample, including their department information, and whether they were theorists or experimentalists.}
    \label{fig:educator}
\end{figure}

The interviews were conducted via Zoom in a conversational manner. All interviews were recorded and transcribed automatically. Transcription errors were corrected by listening to the recordings. Repeated words such as “you know”, “like”, “sort of”, and other similar common filler words/phrases that participants used in conversations were removed from transcriptions for clarity. We conducted thematic analysis for organizing the data, where for the first round of coding, the researchers used structural coding~\cite{saldana2021coding,hedlund2013overview}, which is a holistic approach that organizes responses based on the asked research questions, which we refer to here as our main themes. In the second round of coding, two researchers coded the responses to different questions based upon similar emerging patterns within each theme. These codes were refined through several rounds of brainstorming and discussion, and all researchers agreed with the final coding represented here. Although all 13 quantum educators provided their insight on many different questions, here we only focus on the responses from nine of the educators as their responses conveyed the important ideas, avoided repetition, and maintained clarity. We also note that not all educators were able to answer each question or had varying length of responses. Out of the four educators not included, two educators noted that they did not have explicit thoughts related to RQ1 and the other two educators did not have the time to address RQ1. Similarly, all four of the educators not included did not have time to address RQ2. Also, all of these educator's responses to RQ3 were similar to the responses reported below. Therefore, we found it appropriate to not include these educators for conciseness and clarity purposes. Information about the nine educators included in this paper can be found in Figure \ref{fig:educator}.

\section{Results}

\begin{figure}
    \centering
    \includegraphics[width=0.85\linewidth]{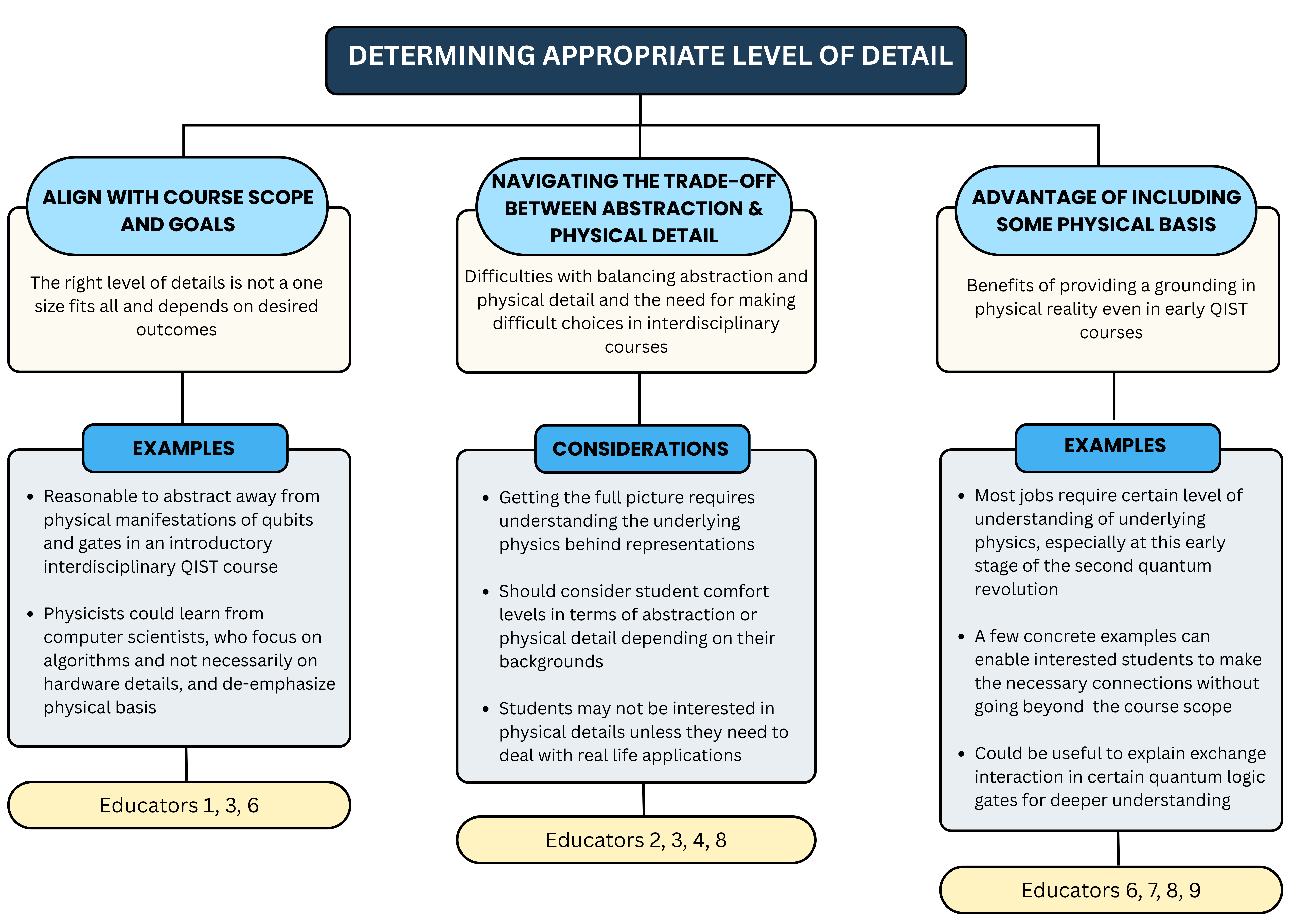}
    \caption{Common codes and descriptions of educator responses to determining appropriate levels of physical detail vs. abstraction in their instructional approach.}
    \label{fig:RQ2}
\end{figure}

\begin{figure}
    \centering
    \includegraphics[width=\linewidth]{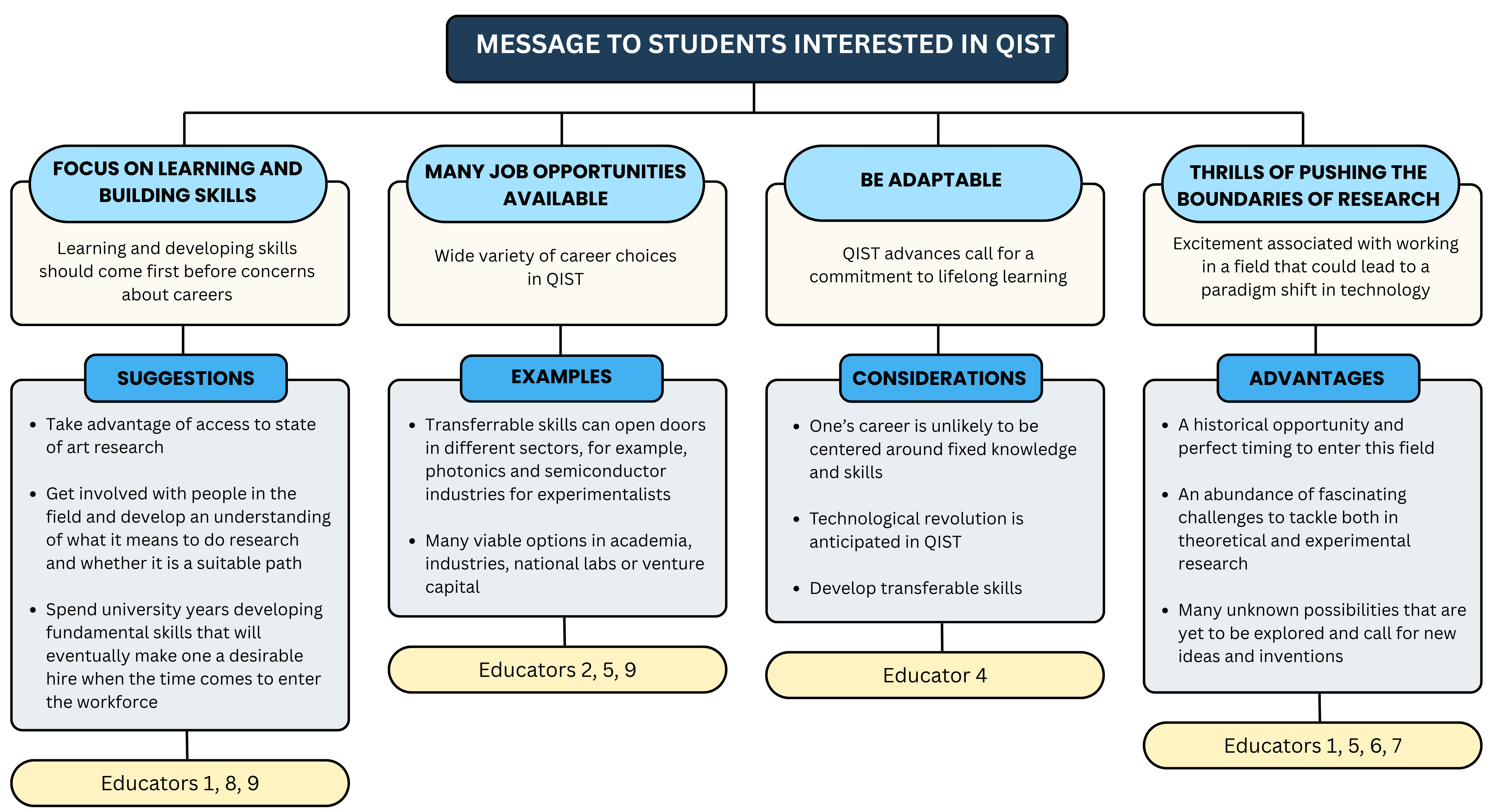}
    \caption{Common codes and descriptions of educator responses to messages to students interested in QIST.}
    \label{fig:RQ3}
\end{figure}

Below, we discuss our findings for each of the research questions. The results are organized by the recurring codes that emerged as patterns in responses, grouped within the broader structural categories/themes pivoting on the research questions. Figures \ref{fig:RQ1}, \ref{fig:RQ2}, and \ref{fig:RQ3} provide a summary of findings for each of the research questions, in addition to information about which educators discussed those ideas. The results section is then followed by an overall discussion of our findings, summary and conclusions and future directions. 

\subsection{RQ1. Evolution of a common language: How can we develop a common language for teaching interdisciplinary QIST courses that serves students from diverse backgrounds equally well?}

\begin{figure}
    \centering
    \includegraphics[width=\linewidth]{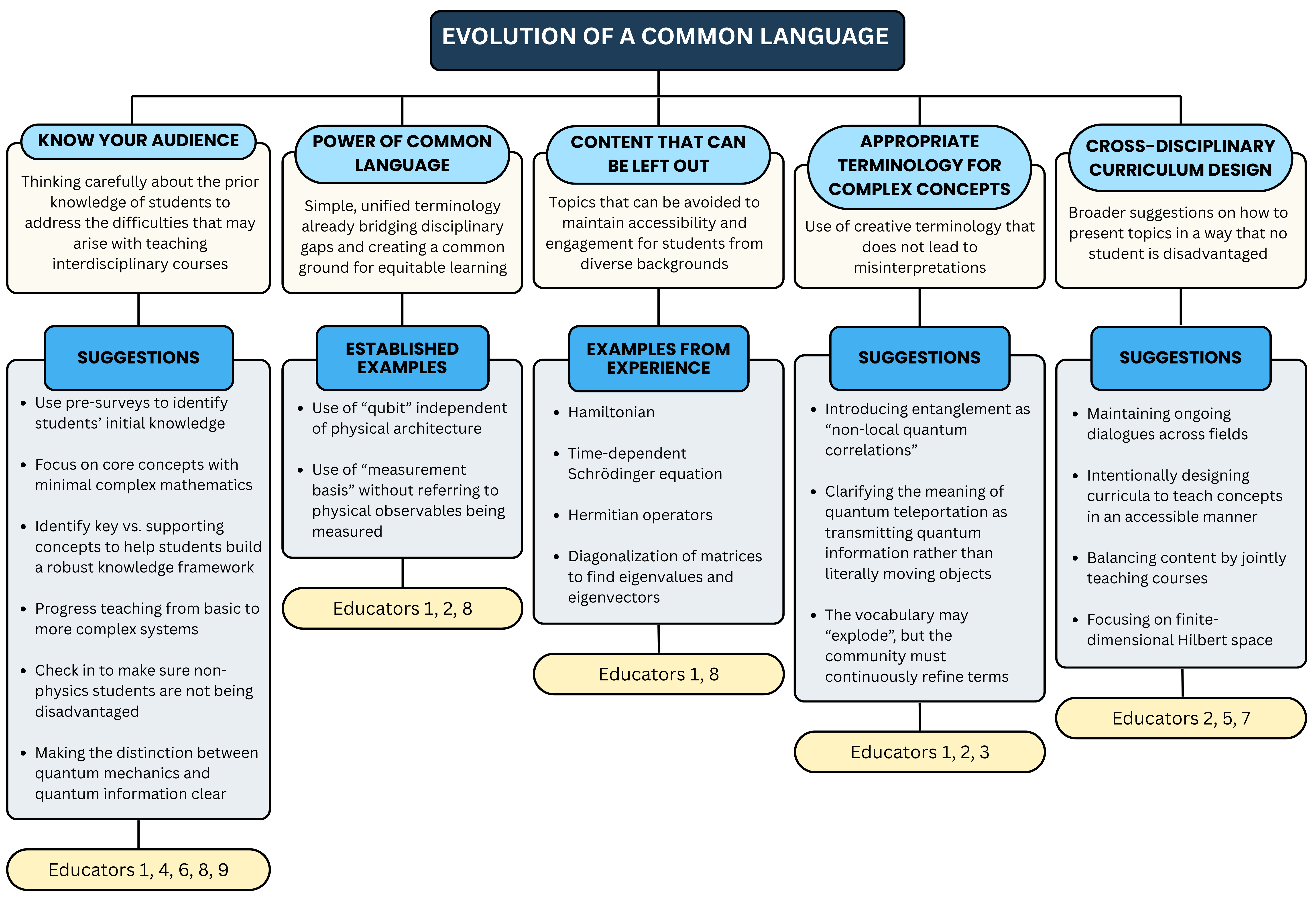}
    \caption{Common codes and descriptions of educator responses to developing a common language for teaching interdisciplinary QIST courses.}
    \label{fig:RQ1}
\end{figure}

The educators had many productive suggestions in response to this research question. They acknowledged that balancing the needs of students from different disciplinary backgrounds in QIST courses at any level is difficult, especially in introductory foundational courses. They also thought that instructors should think carefully about the prior knowledge of their students, goals of their course, and what key concepts they wanted their students to master. They stressed the importance of communicating with educators from other disciplines (e.g., physics educators should talk to engineering, computer science, and chemistry educators) to better understand what topics are appropriate to teach students in a particular interdisciplinary QIST course. They also emphasized that there is definitely power in common language, and coining terms such as ``qubit" and ``measurement basis" has greatly made QIST more accessible to non-physics majors. While the educators stressed that deciding what content is appropriate and what can be omitted is challenging, they believed that instructors should deliberate these issues carefully with educators from other disciplines. They also discussed challenges in creating terminologies for complex QIST concepts such as teleportation and entanglement. The recurring codes for this theme include: ``Know your audience", ``Power of common language", ``Content that can be left out",
``Appropriate terminology for complex concepts", and ``Cross-disciplinary curriculum design", as presented in Figure \ref{fig:RQ1}. Below, we discuss educator reflections on each of these ideas.

\subsubsection{Know Your Audience}

Educators emphasized the necessity of focusing on interdisciplinary students' prior knowledge and skills and building on those in a meaningful way so that non-physics students are not disadvantaged in QIST courses. They also noted deliberating carefully about the key concepts that students in a given interdisciplinary QIST course must learn. 

For example, Educator 1 noted that in interdisciplinary  QIST education, to ensure that all students benefit from instruction regardless of their disciplinary background ``it's always important to know who your audience is, whether you're giving a single lecture or a whole course. Having some understanding of the basis of student understanding, I think, is very helpful." They provided an example on how to reach this understanding saying, ``Having a [pre] survey, for example, that asks students, do they recognize certain physics jargon or quantum computing jargon. And if so, what did they think it means? Having that as a starting point, as feedback, is probably very helpful to any instructor." They continued,  ``I think in designing a course ... when you work with a textbook, whether it's your own or somebody else's, you want to be attuned to how many distinct concepts you're introducing, and make sure that they're sufficiently well connected so that students build a knowledge framework, and not just a rickety set of words and catch phrases. So, I think if you're going to do it yourself, it's a lot of work to make sure that you do it right ... a better way to do it is to understand the structure of what you're trying to teach people, what are the key concepts and what are the supporting concepts, and how do they connect to each other".

Educator 1 reflected further on the key concepts that students should learn in the first interdisciplinary course on foundations of QIST, which is open to all science and engineering majors with only introductory calculus courses as prerequisites. They emphasized the need to focus on fundamentals that all students must learn saying, ``students need to understand the quantum nature of quantum information, they need to understand a qubit backwards and forwards. I think that's probably half of it, or a little bit more than half. Then they need to understand 2 qubits so they need to understand how they can be together, or 3 [qubits] let's say, and that 3 qubits give you 8 possibilities. They need to understand how the thing scales, how to think of it, how you change states, how you represent states, and 
what causes them to change. Then you have this kind of musical score [quantum circuit diagram], which is the flow diagram of a quantum algorithm, a quantum computation, and they need to understand how that works. Then how to initialize gates and readout of a simple system of 2 and 3 qubits. I think that working with the small skeletal structure like that ... would help students get a handle on everything. So, working with small quantum systems I think is very important, making things very concrete for students because this stuff tends to be very abstract. Other concepts like entanglement and other things, those are going to be important as well, because we think that that's an essential part of quantum computing. The power of quantum computing is this idea of entanglement. So, I think that having a small intellectual foundation to define these situations, these concepts, to instantiate them, to make them clear [is important]". 
 
Educator 4 reflected on challenges in finding a common language in interdisciplinary QIST courses saying, 
``That's something I think a lot about, [partly] because of the interdisciplinarity of the field, you absolutely run across it. Every single person hits the limits of what they understand. You need to understand something in depth, but you can’t understand everything in depth. So how do you pick? How do you understand enough and yet not get totally bogged down?... to teach someone else and say, oh, here’s the minimum standard [is challenging].” Yet, Educator 4 thought that finding that minimum standard is crucial.

Educator 6 drew an analogy between teaching a physics course for non-science majors and teaching an introductory, interdisciplinary QIST course. They said, ``well, it's clear that they are different audiences". They reflected back on past courses for non-science majors they had taught saying, ``30 years ago, I thought courses like Physics for Poets were not useful. I thought that was a joke. Like, you can't teach physics without equations. But then as I started teaching courses like that, like the physics of music, and the physics behind the internet for non-scientists, I started to realize that physics is not really mathematics, it's really concepts. It's just that the mathematics is the language that we use to express the concepts very concisely. But the concepts [you need] to explain [should be] with very minimal mathematics." Regarding key quantum concepts, they thought that even though different educators may not agree on what they are for an interdisciplinary first QIST course, ``it is possible to develop a working intuition for how quantum mechanics works at sort of an operational level. We have all these axioms, the postulates, the rules, and so on ... you would try to teach the beginning students those things: that physics is models, it all comes from experiments, it may be counterintuitive, but it corresponds to the real world, because we tested it by experiment. And secondly, the fact that it seems so counterintuitive, yet it still models the real world very well".

Educator 8 discussed their interdisciplinary course which they tried to ensure is suitable for all students: “I just taught a class and I know that not all of them are physicists. So, I try to provide a background that's not the full quantum mechanics we typically teach in physics, but enough that I think [is needed] for the class.” Although polling students in an interdisciplinary QIST course about their backgrounds, including whether they are physics majors or non-physics majors at the beginning of the course, can be valuable for developing a framework for QIST education and for monitoring student progress,
Educator 8 had not tried to poll their 
students' backgrounds.
They reflected, “I just didn't do an analysis to see whether the division [between who was doing well and who was struggling] was along physics and non-physics [majors].” They clarified, “Mostly, I think it's because I didn't actually know who was [from] physics and who wasn't. I knew certain students were doing well, and certain students weren't doing well, but I didn't correlate them to the major they were in.” These reflections suggest that collecting information about their students' backgrounds in an interdisciplinary QIST course is something that can be a helpful step towards monitoring student progress and development of a framework for effective education.

Regarding knowing your audience, Educator 9 noted that they always want to make the distinction between quantum mechanics and quantum information clear in the courses they teach saying, ``I think it's really important, because 100 years ago, quantum mechanics was the revolution. But now the subset of that is really where the revolution is". They felt that de-emphasizing physics ideas would help students from interdisciplinary backgrounds master these concepts well in interdisciplinary QIST courses.

\subsubsection{Power of Common Language}

Several educators reflected on how some concepts, such as the concept of a qubit (as opposed to discussion of specific two states in different physical systems), quantum logic gates (as opposed to explicit discussion of time-evolution via Time-dependent Schr\"{o}dinger Equation), quantum circuit diagrams, measurement basis (as opposed to explicit discussion of eigenvalues and eigenstates of Hermitian operators corresponding to observables), etc., provide a common unified language that help avoid overwhelming students from non-physics backgrounds when it comes to discussions of specific physical systems. Even phrases such as ``quantum mechanically" used by physicists for many decades have been abbreviated to ``quantumly" in the context of QIST.

Educator 1 provided a concrete example of how the concept of ``qubit" independent of the physical architecture is unifying the language and is especially beneficial for non-physicists. They said, “I do think that the language of qubits is a very powerful one, and it does have the potential to provide some unity and common language. It certainly helps within the field of quantum information, which may be dominated by physicists. But you have all different types of qubits, and a qubit is really just 2 quantum states out of a much bigger space that you’ve identified, where the dynamics is mostly contained between those 2 states and [there is] very little interaction with the rest of the world, and you have the ability to actually control those interactions and couple them to other qubits.” 

Educator 1 stressed that talking about qubits as opposed to specific two-state systems such as spin-1/2 or photon polarization states can provide clarity to students from interdisciplinary backgrounds. ``The most trivial quantum system is that in which there's only one state; we don't talk about that because it doesn't do anything, just acquires a phase, and there's no way to distinguish it from any other state. But 2 [two quantum states or a qubit] becomes incredibly rich. Even when we talk about the most complicated materials systems like superconductors with Avogadro's numbers of electrons and interactions, we can still define 2 quantum states [qubit] and talk about them and exploit those [two states]. That's the elegance of having a [qubit] subspace in an incredibly high dimensional Hilbert space and then abstracting it down to 2 [states]". They also emphasized that the reason they find ``qubit" to be a powerful common language for QIST is that “in any 2-level system, the techniques are just abstracted away from the physical origin. It doesn't really matter which two states you’re talking about, whether they’re hyperfine levels in some atom, or whether there’s some ground state and excited state of some quantum dot or whatever. The point is that you can map what you do from one realm into another realm.” 

Educator 1 also stressed that “there is a common language about, for example, electron spin resonance or spin resonance techniques that were first developed in the context of nuclear magnetic resonance” that help provide unification and can help students from interdisciplinary backgrounds learn QIST concepts.

Educator 2 similarly focused on the word ``qubit" and how in the early nineties, it helped pave the way for a common language. They noted that the word qubit came into existence by recognizing that a qubit can be thought of as a unit of quantum information saying, “Another huge thing … was Ben Schumacher’s noiseless coding theorem \cite{schumacher}, which actually introduced the word qubit,” as a unit of quantum information. “This was Ben’s question: if you have a whole bunch of quantum states, say, photon polarizations, for example, could it be that you can compress the information contained in these states into a smaller number of objects? Say, compress that quantum information into a smaller number of photons? And he proved that yes, you can do that, and then you can do a decompression so as to recover the original information. This is Ben Schumacher’s noiseless quantum coding theorem  \cite{schumacher}. It’s analogous to Shannon’s noiseless coding theorem for classical information … and it introduced the word 'qubit' and really focused attention on the quantum state as a kind of information in its own right.” Educator 2 continued, “I think this theorem, and the word qubit, really helped to define the turning point between what I think of as the old quantum information theory, where everyone was concerned about classical information (which was interesting stuff)” to “more of a focus on the quantum information itself. Quantum teleportation \cite{teleportation,photonic} is part of that transition also, because you’re conveying quantum information from one place to another.”

Expressing views on a common language further, Educator 2 reflected on the emphasis on the ``measurement basis" in quantum information as opposed to physical observables saying, “in the use of the word `measurement', there’s a change of emphasis that has happened in quantum information theory, as opposed to earlier quantum writings and discussions. In standard treatments of quantum mechanics, a measurement is described mathematically by an operator, an observable, which has eigenvectors and eigenvalues. From the quantum information perspective, what’s important about a measurement is almost always the eigenvectors and hardly ever the eigenvalues. So the focus has moved into, what is the basis that you’re measuring in? Now, of course you do it physically, so there’s some physical process that you do to make a measurement. For example, in the case of the Wollaston prism and the 2 photon detectors, you're measuring polarization in a particular basis. I really don’t care what observable we associate with that measurement. I don’t care what eigenvalues we associate with horizontal and vertical polarization … you can call them 1 and 0, it doesn’t matter to me. All that matters are what the 2 vectors are that you’re distinguishing. So, this is something that’s different [about QIST compared to standard expositions of quantum mechanics]." They continued,  ``I’ve encountered this difference in giving talks to a physics audience: maybe some of the audience members, when they think about a quantum measurement, they’re thinking about some observable being measured. That’s really not the way quantum information people typically think about a quantum measurement. The eigenvectors of the observable are what you're paying attention to … So that’s one difference in language
that I’ve adopted.” However, they noted, “I don’t use the word ‘quantumly’.”

Common language does not only apply to physics terminology, but it also pertains to the level of mathematical details for students from interdisciplinary backgrounds, who may not have had certain prerequisites. To make sure non-physics majors are not disadvantaged and find a common ground, Educator 8 reflected on how they have kept accessibility in mind in their first QIST course, as some students may have been learning quantum mechanics for the first time saying 
``I basically structured this so that [I consider] what's the minimum amount of basic stuff I can do so I can get to the next topic”.
Educator 8 kept the mathematical challenges in this course with minimal prerequisites to a level that most students from different backgrounds would be able to tackle the expected types of problems in the course. They said that although many students had not taken linear algebra, they were able to do the basics they introduced, “I needed minimum amount of linear algebra. All I needed was basis transformation and matrix multiplication. I don't think technically that [linear algebra] is a prerequisite, but almost everyone who took that class knew enough so that I didn't have trouble with that.” Educator 8 also said, ``physics majors [in my course must have] benefited from already having a good, solid foundation" in traditional quantum concepts, but they felt that their course provided adequate opportunities for learning foundational QIST concepts, e.g., quantum algorithms, that are not covered in quantum physics courses saying, “a lot of algorithms [I discussed] built on a lot of other topics beforehand, which includes states, entanglements, and multi-qubits".

\subsubsection{Content That Can Be Left Out}

Some educators noted that one should explicitly think about what quantum physics concepts are not necessary at least for a first QIST course on foundations. They noted that they would mainly focus, e.g., on quantum logic gates (and circuit diagrams) and would not discuss the Hamiltonian for a quantum system and how it governs its time-evolution via the time-dependent Schr\"{o}dinger equation. They also said that they would not discuss physical observables and how for every observable, there is a Hermitian operator. Since they would avoid these discussions about observables and Hermitian operators associated with them, they would also not need to teach their students to diagonalize matrices corresponding to different Hermitian operators and find their eigenvalues and eigenvectors. 

For example, Educator 1 emphasized that those teaching interdisciplinary QIST courses on foundations should balance student needs based upon their diverse backgrounds, and not only come up with a common language that resonates with a majority of students, but also center students’ prior preparation as well as their mathematical backgrounds by thinking about what can be removed from the course especially to benefit students majoring in disciplines different from physics. They said, “think about avoiding parts of the theory that are overly complicated but aren't necessarily essential for what you want to communicate or describe”. They felt that instructors teaching interdisciplinary courses with non-physics majors in them should recognize that even though many [non-physics] students may be interested in ``the theory” underlying the concepts covered, they may only be interested in ``rate equations, for example, or something that can be described without really writing down all of the quantum states and [they may not be interested in] thinking about matrix elements and Fermi's Golden rule \cite{griffiths}” that a quantum physics instructor would typically delve into in great detail in a course for physics majors.

Educator 1 noted that if educators think carefully about how to connect meaningfully with students at a given level, it is possible to teach foundational QIST concepts to an interdisciplinary audience at both high school and college level; even if they do not have any prior knowledge of quantum physics or significant mathematical preparation. They said, ``there are ways of abstracting away some of the more challenging mathematical prerequisites, to just integrate them out, for example, the [time-dependent] Schr\"{o}dinger equation [for time-evolution of the system]. It's not like you're getting rid of the Schr\"{o}dinger equation; you're just integrating it [when you focus, for example, on quantum logic gates, which are solutions of the Schr\"{o}dinger equation over a finite span of time]. This is like the integral form of the Schr\"{o}dinger equation, which is a unitary operation". Educator 1 stressed that it is these unitary operators that make up quantum logic gates that are used for time-evolution in quantum circuits, e.g., in quantum computers, and it is good to omit the time-dependent Schr\"{o}dinger equation in an interdisciplinary course on foundations of QIST so that non-physics majors do not get bogged down in the physics involved.

As noted earlier, Educator 1 valued the power of the unifying ``qubit" concept. Regarding teaching about different qubit architectures (e.g., superconducting or semiconducting qubits, neutral atoms, etc.), Educator 1 felt that they did not see the necessity to discuss different architectures in detail while teaching QIST concepts to students who are learning them for the first time. They said, "I don't think it's necessary honestly to talk about spins and superconductors and all of this. I think they should know that it's out there. If you're a physicist you might explore this, but I don't have any moral qualms as a physicist in collapsing that [to simply qubits, gates, quantum circuit diagrams, quantum algorithms] and not worrying about it [physical architectures] because we don't have to. There's so much to talk about without going into those details. And I'm saying that as somebody who does experiments and works with the physical and materials aspects of quantum information. For this type of [first] course [in QIST], I don't think it's necessary". However, Educator 1 felt that a short discussion about them may be useful saying, "I might spend a little bit of time just for cultural reasons so that they understand that it [different physical representations of the qubits using different architectures] is out there. But I wouldn't spend a week on it". 

Educator 8 reflected on why they did not focus on the diagonalization of matrices corresponding to Hermitian operators for finding the eigenvalues and eigenvectors in their interdisciplinary QIST course saying, “So, for the sake of time and accessibility, I basically did not discuss eigenvalues and eigenvectors, because while they're necessary for certain physics problems, working off Thomas Wong's book \cite{wong}, I mostly managed to basically avoid them.” They explained why they did not talk about measuring physical observables and eigenvalues of Hermitian operators corresponding to physical observables saying, “I only did measurement in the computational basis and because I did the gate-based method, you can always transform any other measurement in a different basis into that one, and so it was sufficient. Again, for the purposes of the class, [it was sufficient] to just talk about measurement in the computational basis”. They emphasized that they wanted to prioritize other topics for this course, ``I figured out that I did not need those topics [eigenvalues and eigenvectors] to talk about quantum algorithms and quantum communications and other things [I wanted to discuss]."

Educator 8 also noted that they did not discuss the Hamiltonian of a quantum system saying, "if I were to talk about Hamiltonian, it would be probably more in the context of physical implementations. From a purely information [theoretic] point of view, I could skip the Hamiltonian”. They added, “for the most part…I didn't get too many questions, [it was not the case] that [students] really wanted to know about these details of how does the Hamiltonian work? Occasionally, I did get a question about, oh, how would you implement this gate and stuff like that and…I would probably have to go a little bit into Hamiltonian physics for that to work…but…for the most part, I didn't get that many questions in that area.”  
Educator 8 thought that avoiding discussion of the Hamiltonian helped ensure that physics majors did not have an advantage in that case. They felt that their choice of skipping the discussion of the Hamiltonian was consistent with the fact that they focused on gate-based quantum computing to keep the discussions at a level where all students regardless of their disciplinary backgrounds can understand them. They said, “It was a deliberate choice not to talk about the Hamiltonian because, again, this is following Thomas Wong’s book \cite{wong}. He does gate-based quantum computing, for which you can avoid a Hamiltonian language". 

\subsubsection{Appropriate Terminology for Complex Concepts}

Some educators discussed how creating appropriate terminology for complex QIST concepts such as quantum entanglement or teleportation is challenging, yet deep consideration of terminology is very important for helping students understand these concepts well. They stressed the challenges in coming up with suitable terminology and language to communicate about these interdisciplinary QIST concepts in ways that they are not misunderstood or misinterpreted, especially by beginning students.

For example, Educator 1 reflected on the challenging nature of teaching the concept of entanglement saying, ``I would say that a concept like entanglement isn't the first or second thing that you talk about. You need to get the basics down". They elaborated further on why quantum entanglement can be difficult to teach despite this term possibly evoking some productive ideas, saying other concepts such as qubits and quantum measurement must be understood before learning about entanglement. While they emphasized the importance of language and terminology, they also stressed that instructors should make sure to help students learn the prerequisites; e.g., for entanglement, measurement is a prerequisite. They said, ``Entanglement- it's actually the words that we attach to mathematical or physical concepts, that have a big influence on how we think about them. Maybe that's a good thing, but sometimes it's bad, because if you use a word that doesn't evoke the physics, or it is inconsistent with the physics or the mathematics, then you run into lots of problems. I would say that entanglement is somewhere in between. There are parts of the word that are good in the sense of being tangled up. If I think about a cable that's tied in knots, it's all tangled, and you pull on one end, the other one goes with you, [and] pull the other one, and the other first half goes. So, their outcomes, their fates, their motions, are correlated. So, you could say, well, there are correlations there, correlations between 2 qubits, or you can say they're entangled ... That kind of physicsy, visual representation of the word entanglement is not bad, but without a thorough understanding of what quantum measurement is, and what qubits are…I think it's very hard to truly understand what entanglement means, even in the case of 2 qubits; because we can think about classical versions of entanglement or correlations that are not really the same as the quantum mechanical ones". They thought that QIST instructors should make sure students learn that entanglement is ``non-local quantum correlations" and use this concept in concrete operational cases to help students ``figure out how to make a quantum key distribution protocol" and "learn how to do distributed entanglement or remote gates and quantum computers, etc." They felt that ``if the students can actively and truly understand what the weirdness is and actually answer questions about how these quantum correlations take place", they would have a better understanding of what quantum entanglement is than by just learning about it abstractly.

Since Educator 2 played a key role in formulating the concept and coining the phrase ``quantum teleportation" \cite{teleportation}, they said that they understand how the word ``teleportation" can cause confusion. They acknowledged that since this type of language can be misinterpreted, it is critical to emphasize to students that quantum teleportation moves a ``quantum state" and not physical entities from one place to another. The historical account of Educator 2 about how the quantum teleportation concept was formulated and this concept and terminology came about is included in the Appendix (this valuable information is not available elsewhere). 

Educator 3 reflected on the challenges in creating new terminology and language for QIST saying, “I think it’s probably going to get way messier. Well, if we’re all being very creative, inventing all these new ideas and new ways of thinking about stuff, then the language is going to explode and get very confusing. And only then, once we kind of understand it, can we trim it and clean it up. I don't think you develop a field with a crisp, well-defined language at every point in time. I think, especially if your field is going broadly, you have an explosion of language. Then you sort of define it and clean it up and textbookize it later.”  They also acknowledged, “for instance, when I say, measure the qubit”, then “there it’s implicit that I’m measuring in the computational basis of the qubit so $\sigma_z$”.

Educator 3 also gave specific examples of interesting language they had recently heard saying, “there’s a fun term called magic in gates related to their ability to generate entanglement like this gate has magic, or this gate doesn't have magic. On the one hand, it's confusing. On the other hand, it’s someone earnestly trying to describe to you some abstract property of these gates that they're studying. So, I actually kinda like watching the language explode. I agree it's messy. I agree you have to keep redefining terms.
But the fact that the language is exploding means that we’re doing lots of interesting things and thinking in new ways. So, I think that part is good. And maybe it’s too cynical of me, but I feel like it won't really get regularized until most of the excitement is over.”

\subsubsection{Cross-Disciplinary Curriculum Design}

Educators also had some broader suggestions about cross-disciplinary curriculum design for QIST \cite{fargol} and how to help students learn these exciting concepts in a manner that non-physics majors are not disadvantaged.

Educator 2 reflected on a course they had taught multiple times jointly with a math professor on protecting information for roughly 50 undergraduates majoring in physics, mathematics or computer science, which they believed had a good balance of content to equally challenge both physics majors and non-physics majors. They explained that physics majors had the advantage of prior physics knowledge, whereas math majors were good at proving theorems and computer science majors were good at algorithms. They believed this contributed to the success of the course and hence, they have offered it multiple times. They also noted that this course was not specifically easier for any particular kind of student, but it certainly helped to have a science background saying, ``to [be] one of those 3 majors was very helpful…but we had sophomores, juniors, and seniors in the class.” They also noted that ``we reminded people about how linear algebra goes…but it certainly helped to have seen it before ... we certainly spent some time going over complex numbers ... because not all the students had worked with complex numbers before”. They also added that in their interdisciplinary course, ``there were challenges for everybody. Yes, the math students were not familiar with the physics. The physics majors may or may not have been used to proving theorems ... I don't think any particular kind of student had a strong advantage.”

Educator 5 reflected on interdisciplinary issues in QIST education and highlighted the critical roles of discussing these issues with colleagues in engineering and computer science, reducing mathematical challenges and focusing primarily on finite dimensional Hilbert space. Stressing the importance of focusing on finite dimensional Hilbert space in QIST context they said, ``if you look at, for example, the introduction to quantum information, the textbook by Nielsen and Chuang \cite{nielsen2010quantum}, and of course we're going back 20 years in these things [since the book was written two decades ago], but already they had I think 95\% of the right ideas that you could take from quantum mechanics [for physicists], the elements of quantum mechanics that you really needed to understand quantum information science and boil it down to a relatively small set of mathematical operations. It was based around linear algebra, and so solving complicated differential equations was not an issue ... [the authors emphasized] dealing with these things as linear algebra problems and finite dimensional Hilbert spaces". 

Educator 7 had helped establish a quantum engineering bachelor's program at their university (details can be found in \cite{fargol}). The program was intentionally designed to present key quantum concepts in language accessible to non-physicists, as its courses are primarily geared toward engineers. The educator noted that their focus on concrete applications in the courses in this program is also beneficial to the few students who have already taken quantum physics courses in the physics department.

\subsection{RQ2. Determining appropriate level of detail: What level of physical detail vs. abstraction is appropriate for an interdisciplinary QIST course for students at a certain level?}

In response to this question, while some of the quantum educators thought that they would definitely teach some physical basis, e.g., of quantum logic gates such as CNOT, even in an interdisciplinary QIST course focusing on fundamentals, they recognized that it is difficult to balance the level of abstraction and discussion of physical basis of various concepts. They emphasized that whether students are able to answer the question (mentioned as an example during the interviews about whether entanglement of qubits requires a direct or indirect interaction between them) correctly or not depends on the scope and goals of the course. They also noted that students within a given course who were from different backgrounds may have different expectations for whether the physical basis of QIST concepts such as quantum logic gates should be discussed. They noted that instructors should carefully deliberate and decide the appropriate balance in a given course at a certain level and how to connect the level of exposition to the prior knowledge of all students regardless of their disciplinary background. The recurring codes for this theme are ``Align with course scope and goals", ``Navigating the trade-off between abstraction \& detail" and ``Advantage of including some physical basis". 

\subsubsection{Align with Course Scope and Goals}

Some educators emphasized that the answer to this question depends on the goals and scope of the course, as interdisciplinary QIST courses could be targeted to students at different levels even if they are undergraduate courses. For example, the scope of a QIST course focused on foundations may vary by whether it does or does not have prerequisites such as linear algebra, introductory physics, modern physics, etc.

For example, Educator 1 noted that while we may want students to understand that the CNOT gate has a physical origin and requires direct or indirect interaction between qubits, we should also keep in mind the appropriate level of abstraction for beginning students who have not taken many prerequisites in an interdisciplinary QIST course, because the scope of the course is an important consideration. They said, ``I think as a physicist we know that things like CNOT gates and whatnot all have physical origins. Quantum information is physical but if you're teaching a multidisciplinary course where maybe students have never taken a physics course, it's hard to even get them to understand what this question means ... So, what they're doing in saying that ... [we] just [need] CNOT is ... they have honed in on the answer but with the physics scraped out of it ... A related question is, is it necessary to teach them the physical origins of things and even that I'm a little bit ambivalent [about in a course without many prerequisites]. I think it's maybe important to explain to students that these things do have their basis in physics. But to say that we're not going to talk about the physics [or] to say that all of these things, all of these gates and everything, they all have manifestations, which depend on the type of qubit and the physical platform, this [level of physical detail] rests upon [the goals of the course]". 

Educator 1 continued, ``I think they should understand that there is a kind of mapping between [quantum logic gates like CNOT and their physical implementations, e.g., through interactions]. 
Let's say there are 5 different approaches to building a quantum computer. Well, making a 1 qubit gate and making a 2-qubit gate can be completely different processes. But the circuit representation of what they do can be the same in the end and there's something very elegant about that. I don't think that it is necessarily bad; the fact that they are giving a distilled or an abstracted version of this. It is what we aspire to, I think, in this field of quantum information that ultimately, we will be able to successfully abstract away from the physical manifestations, that it won't matter whether it's a superconductor or a spin [qubit], or whatever [other qubit architecture we have] ... In the same way that when we think about [classical] computing, we don't think about what kind of transistors we have inside of our personal computers. In some [broad] sense it doesn't matter. But of course, it does matter because without that the computer wouldn't function. So, if you're teaching a course in the physical foundations of [classical] computers, then, yeah, you should talk about transistors, PN junctions and field effect transistors. But if you're talking about algorithms or computer architectures, it's not relevant. So, I think it depends on the scope of the course [what kind of answers students should be expected to provide for these types of questions]".

Reflecting on the example provided for this interview question, Educator 1 thought that it is possible that if the question was posed differently, students may recognize that direct or indirect physical interactions are needed to entangle two qubits. They reflected, ``Maybe to be even more specific, you can say, well, if you start all [qubits] in the 0 state, you start in some unentangled state ... well, is it possible to do local operations to an unentangled state on each of the qubits and then produce an entangled state? The answer is no. So, I think there are lots of ways one could ask these questions, and each one might reveal a different aspect of what students understand and what they don't understand. I think what was meant here in this question was that ... you want to make sure that students understand that you can't produce entanglement, let's say between 2 unentangled qubits without having an interaction that entangles them."

Educator 1 added that they would talk  ``about unitary operators because those are anyway closer to gates in the circuit representation of quantum computing", and that it is ``less intellectually expensive to talk about unitary operators". They added, ``It's all about revealing the physics origins of all these gates. So, if I think about a spin as a qubit, then a spin interacting with a magnetic field, that's a 1 qubit gate, 2 spins interacting with each other, that's a 2-qubit gate. But those operations can be written as unitary. You can basically write out unitary operators to represent the interactions, basically integrating the Hamiltonian over a finite interval of time that gives rise to a unitary gate and that could be square root of swap or it could be a swap operation or it can be single qubit rotations. All of those things are the building blocks of all quantum algorithms". However, despite emphasizing that the scope of the QIST course is an important consideration, they thought that it may be valuable to allocate a small amount of class time even in the first interdisciplinary QIST course with very few prerequisites to the physical origin of quantum gates saying, ``I might spend a little bit of time telling people that there's something underneath all of this and that it's related to physics. I would mention that the things that produce those circuit elements can be very different, depending on the physical basis of what's going on. But in the end, what matters is what the circuit is calculating".

Educator 1 also noted that offering students collaborative course projects with the option to select projects varying in their level of physical detail could provide students from diverse academic backgrounds the chance to engage with these topics at a depth aligned with their individual interests.

Educator 3 also believed that the level of abstraction vs. physical detail depends on the goals of the students taking the course. They said, “I guess it depends on who they are. Like, if you’re just using the computer, you want to abstract all that away. You just want to say, here’s my algorithm, and it compiles and runs.  You don’t care which qubit is physically connected to what.  If you’re someone that’s messing with that computer, that’s designing computers, you absolutely need to understand it. So yeah, definitely in my quantum class, we did a fair bit of, here’s an interaction. Here’s how it makes a unitary gate. And then we try to talk about a lot of what building a quantum computer is, building different interactions that can make different unitary gates...”

Educator 3 also reflected on computer codes pointing out how those at different levels of the stack may need to know different levels of hardware details saying, “When I use Python, I don’t know almost anything about the level 2 cache. Will this code interact with the level 2 cache of the CPU? How much RAM does it take? I find it amazing that my students can write Python commands, and they don't even know what data type they're storing stuff in, is the computer using float or is it using single double integer? I guess to answer your question, we have to define how high up ... the stack they are. At the very top of the stack we don’t want them knowing about this stuff, we just want them using it, and the compiler is designed to make good choices. If they’re supposed to work at the middle stack or down, this is their jam. So clearly, they need to interact with it.”

Educator 6 reflected on the example problem provided saying, “hmm, I think that’s a very hard question. I’ve never tried to do that myself. When I teach physics [to students with non-physics backgrounds], I just try to teach the physics, but maybe with less mathematics. I know some people do believe that a computer scientist could learn the rules without understanding what's under the hood and what the deeper mathematics is". They acknowledged, ``Of course, they’ll be restrained or constrained in what kind of work they can do. But if they’re just trying to apply or develop standard software, it’s kind of like a hierarchy of compilers. I can write Fortran code, but I don't know what’s under it, but just by experience, I do these things, and it does what I want. So I'm sure there's a role for that [in certain types of courses]".

\subsubsection{Navigating the Trade-off between Abstraction \& Detail}

Quantum educators in general agreed that it is not easy to balance the level of abstractness vs. the physical basis of QIST concepts in interdisciplinary courses at any level, especially in a foundations course. They emphasized that educators often have to make difficult choices about what is appropriate for their group of interdisciplinary student body in a given course with certain prerequisites, given the likelihood that they are interested in learning quantum concepts at different levels of the stack.

For example, Educator 2 noted that there is so much “to talk about with quantum information science because there's the physics and there are the algorithms", therefore needing simplifications makes sense. They continued,  
``To get the whole picture you need all those things. Maybe the best way to do it is to have a course where you really do treat each one of those aspects. Say, remember, we’re talking about CNOT gates, and we think of a CNOT as this 4 by 4 matrix. But remember that that matrix represents some actual physical process and it may not be trivial to carry out that process in real life. What it is depends entirely on what kind of qubit you’re using, right?”

Educator 3, who noted that the answer to this question depends on the scope of the course, further reflected on how they would adjust the level of detail based on what students are looking for. They said, 
``As a physicist, I always want them [the students] to understand the physics. It’s kind of like a tautology. It's a question of what is the machine [quantum computer] to them? I guess for [students in the foundations of quantum computing and quantum information course], it should be a little more direct knowledge, because it’s not how to compile and how to run stuff on a quantum computer. It's like, what is quantum? So I guess in such a course, I would try to teach them how you physically build gates. I think it makes the machines [quantum computers] make more sense. But if it was a course on applications of quantum algorithms, I don’t know how many courses there are like that, then it makes sense that they would abstract it away because they got different problems to deal with in that course, like complexity theory and scaling and all of this stuff, and they don’t have room for it.”

Educator 4 felt that those from different disciplinary backgrounds will have different levels of comfort with abstraction vs. the physical basis of quantum concepts. They reflected broadly on how it is typical in many disciplines not to get into the details of how things work and stay at a higher level of abstraction saying, “[This] is a very computer science sort of idea. Oh, look! There's assembler, and then you wanna abstract it up so that you don’t actually have to think about the registers. And there’s an assembly language. Then from the assembly languages, again you abstract things up and then you talk at that level". They continued, ``It's like when you drive your car, you don’t actually think about where the pressure in the fuel injector or whatever is, right? Things get bundled into things such that you push the gas, and you do whatever. In fields like computer science, that’s absolutely bedrock to the field. They are used to thinking in this sort of abstracted way and it's not wrong. It’s just totally incomplete. But again, this idea-it’s a Turing machine. We have them [computer scientists] saying if they can simulate a Turing machine, it can do anything. Then you [as a physicist], are like, well, it doesn't really matter what it’s made out of or what it’s actually doing? So they [computer scientists] are very comfortable with that kind of abstraction, which I think is less so for the physicists, who are less comfortable with it.” 

Educator 4 also noted that many people may not be interested in the physical basis of things unless they run into concrete real life situations that motivate them to want to understand it in-depth. For example, they reflected on how a lack of deep, nuanced understanding can make people confuse parallel computing on classical computers with the workings of a quantum computer. They explained, “They don’t really understand the nuts and bolts. That was my point about the cars, right? If my car stopped working, it’s not like I could open the hood and do anything to figure out why it wasn’t working. And yet, we’re completely comfortable driving around ... So the fact that we don't understand how it’s actually working doesn't bother us at all, more or less, unless it actually breaks. Then you're like, Oh, gee! I wish I knew what was wrong! But you are perfectly happy to use something that you have no idea how it works. It's the same [in this case related to not understanding the physical basis of quantum gates].” They also noted that while it is difficult to explain, e.g., how quantum computers and parallel computing on classical computers are different, it will be easier to explain the difference based upon their applications once quantum computers become useful and they can do certain things and not other things. 
They believed that is when students will start to comprehend this difference,
``then you might be able to get a little farther [in helping them understand why].”

Educator 8 reflected on the example problem provided, noting that even in their foundations of QIST class, “there was an insufficient explaining that a gate requires a type of interaction." They continued, ``Now, in my class, I didn't talk about interaction very much ... maybe they have misconstrued what that term [interaction] really means because I didn't do Hamiltonian". They further added, “Yeah, so that is something that was quite emphasized from the perspective of quantum information, especially the gate-based one about how to generate entanglement. What I did not do in my class and [if my students were given this question] this could be reflected in [their performance] ... how do you implement a CNOT [via direct or indirect interactions], which was something that I didn't really do in detail.” Educator 8 also said, “I did talk about qubit implementations, but we didn't go into details of saying that these are Rabi oscillations, and you can see the lifetime, how to get a lifetime out of that. So these are certain things from physics, what I would say [are] more experimentally relevant things, which I didn't do.”

\subsubsection{Advantage of Including Some Physical Basis}

Some educators felt that even though the answer to this question depends on the goals and scope of the course, and balancing the physical details vs. abstraction generally remains difficult even in a QIST foundations course with varying prerequisites, it is important to have students learn at least a little bit about the physical basis of concepts such as quantum logic gates or qubits, etc., especially at this early time in the second quantum revolution.

For example, Educator 6 noted that it may be appropriate to skip the physical basis in some specialized QIST courses for students with a certain background. However, regarding foundational QIST courses geared towards students from interdisciplinary backgrounds, they did not think discussion of physical basis should be omitted  saying, ``I’m not sure if it has a very practical role at this point because we don't have machines that can compile quantum algorithms. So, what are they going to do with it [if no physical basis is taught]?" Educator 6 continued, ``At this point they probably need to know enough quantum mechanics, at least at this engineering level so they can think about that at the device level to some extent. If somebody gets the degree now in quantum programming, there are companies out there ... But those people [employed in those companies typically] all have PhDs in physics as far as I know, or in chemistry, because I think most of the applications right now are in science. They’re not in finance or databases so I think it'll be another 10 years before just algorithm-based undergraduate degrees will actually be useful". Educator 6 made an analogy with classical computing for computer scientists saying, `` You could ask the same question about classical computing. A lot of computer scientists when they're students, never learn anything about the hardware layer. They don’t know about a flip-flop and a semiconductor logic gate, or anything about those. They see that as a detail, they just think about the software level. So, I think it would be the same in quantum."  They continued, ``You could teach people that [CNOT can entangle two qubits] without them having to know anything about the physical layer, but 
 ... I don't think there are any jobs out there that are going to hire them. Yeah, maybe it would make sense to separate the software from the physical layer and make that separation explicit. So don't confuse them about what layer you’re talking about and teach one or the other.” They further added, “then the sophisticated ones can think about the connections between the two layers, but not every student will be able to do that.”

Educator 7 thought that some physical basis, e.g., underlying the quantum logic gates is beneficial and described the approach they use saying, “I teach them the physical basis, the actual physical and engineering basis of these operations. So, whenever they simulate a CNOT gate between 2 spins in quantum dots, they understand there is an exchange interaction between them, and they understand where the exchange interaction comes from. It's from the overlap of the wave functions. I teach them the symmetrization postulate. I show them the power of the exclusion principle. They actually understand microscopically where those physical interactions come from and then they understand how it manifests in Hamiltonians which have the properties that they can sustain entangling operations. Then, you show them how to simulate that entangling operation and also how to visualize it and quantify the entanglement. It sounds harder than it is.”

Although Educator 8 acknowledged that abstraction vs. physical basis of quantum concepts are hard to balance in QIST courses on foundations, they believed, “It's important to understand the basis of interaction and I think that's an important detail to know." They explicitly referred to the example problem about the entangling gate provided saying, ``I will also admit, had you given this question to my class, there was a good chance they wouldn't do very well on it. One thing I didn't do a very good job on either, some of the kind of the physical aspects, some of the language, like lifetime. I mean, I mentioned what lifetime is, but specifically how you measure lifetime, what it means for two qubits to interact, some of this stuff, that is an aspect that I did not talk about very much.” They added, “I was just kind of maybe doing more of a computer science slant and so it just ended up that way that, in retrospect, I should have covered certain things more.” 

Regarding their undergraduate course on QIST foundations, Educator 9 noted, "I haven't had that come up. I don't know whether they don't think that way or ... [if] my way of lecturing prevents that error. But what I do focus on when talking about entanglement is that it is a correlation and that is very important".

\subsection{RQ3. Experts' messages to students interested in QIST: What guidance should we provide to students considering courses, degrees, and careers in this rapidly evolving field?}

Overall, all educators had very positive messages for why students should pursue courses, degrees, and careers in QIST. Some of them primarily focused on the importance of learning QIST concepts and developing transferable skills. Others focused on how the variety of careers available in this area allows students to find a job consistent with their interest and commensurate with their skills. Some educators stressed the importance of adaptability, especially in careers involving quantum technologies in which the landscape is evolving quickly. Additionally, quantum educators stressed how engaging with QIST is particularly exciting because participating in quantum technology research can give them an opportunity to push the boundaries of research and be part of creating transformative possibilities that we have yet to imagine. Overall, the recurring codes for this theme included ``Focus on learning and building skills", ``Many job opportunities available", ``Be adaptable", and ``Thrills of pushing the boundaries of research".

\subsubsection{Focus on Learning and Building Skills}

Some quantum educators emphasized that instead of currently being concerned about careers in this area, students should focus on learning fundamental QIST concepts and building skills, as many jobs will most probably still be available when they finish their education.

Educator 1 taught a first experiences in quantum course that met an hour each week for first [and some second] year science and engineering majors in any discipline, who would like to conduct research with faculty members in quantum areas. They said that their message to students in that course is that at the university, they have ``access to state of the art research, and there are a lot of people at the university ... who are doing quantum research, and they can get involved". They continued, ``they should understand that it's not all about preparing themselves to enter some workforce where somebody just tells them what to do and they know how to turn a knob. It's an opportunity to understand that we don't understand everything, that research is a frontier and that you have to be an explorer. And you have to learn how to tackle the unknown and the ill-defined statements and things that are not so neatly packaged and that the earlier you can do this the better. And don't worry about the workforce. When you're in college, worry about learning and learning how to learn. And I would put aside those jobs; they will be there when they graduate or go on to graduate school. The jobs will be there, but they need to be prepared to meet the challenges of the 21st century".

Educator 8 said that in response to students who ask this question they say, ``this is a cool area, try to do a bit of research and learn about it and see if you like it. Usually, that's kind of my standard response to any student, see if you like it, because we want people in the field who love what they're doing".

Educator 9 pointed to availability of job opportunities, while also emphasizing the importance of growing intellectually by learning as much as possible from various avenues available, regardless of whether they will end up having a career in QIST. They said, ``Quantum computing is such an intellectually stimulating field. While in undergrad or grad [education], gaining those skills, being able to think about such hard problems, [you will develop] a skill set that can be applied to so many other fields. In case this [QIST] either doesn't work out longer term, or it becomes maybe personally not the best fit. [But] even if this doesn't work out [as a career] ... you’re much better off if you learn about quantum computing now when it's in its infancy. And even though you are in some other field [later when you join the workforce], say you're in pharmacology, at some point, quantum computers can help design better drugs". They emphasized that students will be ahead by taking even one course to understand what quantum computing and quantum information is about. They added, ``both the understanding of what quantum computing is about and then also the skill set you need to develop to work with, those two things are transferable to any other field". 

\subsubsection{Many Job Opportunities Available}

Some educators emphasized that there are many jobs available in QIST and students will have a wide variety of choices available based upon their interest.

For example, Educator 2, while pointing out the exciting potential of QIST, noted how a former student of theirs who recently got their PhD in quantum information said that industry jobs were the most viable options available in this field.

Educator 5 stressed that students should definitely pursue certificates and degrees for careers in QIST saying, 
``it’s good for fundamental research and basic interest point of view, [and] also great from a career point of view. It’s been wonderful; I've had 3 people from my research group in the last 12 months who have [graduated from] my group and gone into industry and stayed in this field doing work in quantum computing for different startup companies, they are doing fascinating stuff! There are opportunities in universities, and there are opportunities in industry. And of course, skills you have can always take you off into a wide range of other things as well. But now there’s also real industry opportunities in real quantum technologies. So, I think it's a very exciting time to be studying this area and we're always on the lookout for keen people to come and join the teams.” Educator 5 further noted, “the skills you learn in ... this type of area have always opened different things across programming, across mathematics and various applications or engineering tools and technologies, a lot of people in experiments going into laser companies or the photonics industry, semiconductor industry and so on.”

Educator 9 pointed out job opportunities in QIST saying, "Oh, absolutely, there's so much opportunity. I mean realistically, there are just so many job openings whether they're in industry or academia or national labs or venture capital...the whole ecosystem is just growing each year...so that's kind of more practical short term". 

\subsubsection{Be Adaptable}

Some educators emphasized the need for being agile and adaptable, reminding students that 21st century jobs in general and QIST in particular demand skills that evolve quickly and students should be ready to learn new skills as they progress through their careers.

For example, Educator 4 was very enthusiastic about students pursuing courses, degrees, and careers in QIST, but wanted to be upfront about the need for adaptability in the workforce. They said, “I think what's gonna continue to be true is that you can be in a field, it can last a long time, but it’s not gonna be your whole career. So, the question is-is it an interesting direction now? But to be aware that as things evolve, you also have to evolve. And right now, it’s a great field. It’s clear to me that you get skills working in it that will provide that path to adapt your way into the future, even when this crash [of the QIST bubble] that I’m assuming will happen [occurs]... And I think it’s a great field in that way, because it does have a lot of different aspects to it that you can find a niche that really appeals to you and is something that you feel is satisfying…”. 

Educator 4 was reminded of someone they knew in the field of AI who felt disappointed with this need for adaptability, quoting that they said,
``I thought I have my credential, and then I’m set for life". Educator 4 elaborated, ``I don't think that’s gonna be true for anybody. It wasn’t even true for us. But I think it's gonna become less and less true as time goes on. I just think that ... having to evolve is just the way things are gonna be because technology advances, and what is important changes as the technology advances. But I think it [QIST] is a great field in the sense that it has this really interesting fundamental side." Educator 4 emphasized the need for both deep expertise and broad yet practical synthesis in QIST, and how this interdisciplinary field requires bringing together partly-understood pieces to advance.
They said, ``this [QIST field] has a very good balance between those things that you really have to get to the bottom of. But again, there are all these things that you haven’t quite got to the bottom of. But you have to understand enough about them [different pieces] to understand the overall structure of what needs to be done [in QIST]. So yeah, it’s a good field. Not in that sense of it's going to last forever, but in the sense that it’s a field with the right kind of intellectual content where there’s a lot of different aspects to it. And they’re all important and bringing them together is the way to push the field forward. All of those skills can be transferred to whatever turns out to be the thing in 20 years.”

\subsubsection{Thrills of Pushing the Boundaries of Research}

Some educators emphasized that it is great to take QIST courses and pursue careers in this area because it will provide them with opportunities to push the boundaries of research regarding what is possible since we are at the beginning of the second quantum revolution and quantum technologies are advancing rapidly.

For example, Educator 1 noted, ``I do think that it's important to make that effort to share the excitement [with students] if you're working in the field. In some sense, we have a duty to explain to students why we're doing what we're doing". They continued, ``And I think that it's important to try and convey the importance of doing research where you don't always know the outcome, and explain that we're in an era where there's a lot of uncertainty about what we can and cannot do. We believe that there's a firm foundation, foundation that rests upon our most successful theory of the universe; and we're now using what we know about that theory and combining that with what we know about information in developing a new field of quantum information and then trying to see where it takes us. But ... none of this is really certain I would say, there's a lot of quantum uncertainty you could say about the field".

Educator 5 said, “It's a brilliant time to get into this area. It’s a lot of fun. If you’re interested in experiments, there are a whole lot of really interesting challenges that range from fundamental physics to engineering [problems] that need to be solved in every lab. Working in quantum technologies as a theorist, there are numerics to be done. There’s really fundamental analytical stuff that's not well understood about why these things work and where these things work and there is so much to be done that research in this area is really exciting. It’s also exciting because it is being connected to industry. It is being connected to other scientific disciplines. It is being connected to the real world. So, if you want to do physics, learn about what people are doing in finance, or in engineering or computer science, or in chemistry or materials; this is your opportunity to get into a field that is starting to sit at the juncture of so many different areas."

Educator 6 was very enthusiastic about students pursuing courses, degrees and careers in this area and said, ``I would tell them that quantum mechanics was discovered 100 years ago. Initially, after 50 years, it ... led to computers and cell phones and it also led to atomic clocks, which has led to the GPS system, which you probably use every day. We call all those technologies Quantum 1.0 because they're using certain information or certain knowledge we learned about in quantum mechanics ... but I'd say, since about the 1980s, there's been a new recognition that you can go further, and that is, if you can control atoms and electrons and photons at the single photon and single electron level, then you can actually realize these new effects, like superposition and entanglement, which are not necessary in semiconductor [industry] and cell phones or in digital [classical] computers. The new generation is Quantum 2.0, and over the next 50 years, this will create the next huge revolution in science and technology. So, if you want to be at the cutting edge of this field, it's a good place to be. Also, it's exciting because we don't yet know what it will exactly do or be useful for, so there's a lot of opportunity to explore new ideas and new inventions ... it's almost like the AI revolution, where things percolated along for decades and then eventually everything blew up, and AI is now this incredible thing and we expect quantum technology to blow up in about 10 years, maybe."

Educator 7 reflected on this question with great excitement saying, “Well, my thought on that is that they are in an almost unique position in history, having the ability to partake in the creation of something that doesn't exist and has never existed. If you think about it, there are not that many people who can look at the world around them and say, I’m gonna work on something that has never existed before! You can be a lawyer, you can be a doctor, you can be an economist, you can be a mechanical engineer or an electrical engineer. For the most part, you will be partaking of a discipline, of a field, of activity that is established, where there are things that work and things that don’t work; and you will kind of push the edges of that.” Educator 7 compared this to QIST as an interdisciplinary field, “[here] you will partake in building something that doesn't exist. So, if you’re the kind of person who thinks creating something that doesn’t exist is very exciting, this is the time and the place for you…if you’re afraid of investing in something that may not yield results, then this is not for you.” However, Educator 7 also stated that “even if let’s say, you work 5 years in a quantum computer industry and [realize] this is not going anywhere, you can quit…safe stuff is still there.”  Educator 7 added, “But you have a head start on potentially being part of creating something that doesn't exist if you feel adventurous, and if you feel curious enough to do it.”

\section{Discussion}

The perspectives shared by leading quantum educators reveal both the opportunities and challenges in building an effective educational framework for the rapidly evolving field of quantum information science and technology. Consistent with Vygotsky's zone of proximal development (ZPD) \cite{vygotsky}, the framework for QIST education must take into account the evolution of a unifying conceptual language, the tension between abstraction and physical understanding, as well as the importance of adaptability in a transformative field considering we are at the dawn of the second quantum revolution.

\subsection{The Power of Conceptual Unification}

The emergence of the “qubit” as a fundamental unit of quantum information represents more than just innovation in terminology. It reflects a profound conceptual shift that enables interdisciplinary communication without being bogged down by physical details of two state systems. As Educator 1 articulated, the qubit concept provides a powerful abstraction that transcends the specific physical implementations, whether they involve atomic hyperfine levels, two-states of electron spin, or photon polarization states. This type of abstraction layer provided by terms such as ``qubit" (in which physical details are not explicated) can help keep QIST instruction within the ZPD of students from different disciplinary backgrounds and help them learn and communicate effectively without getting mired in system-specific details. The historical perspective provided by Educator 2 highlights how Schumacher’s introduction of the qubit concept in the context of his noiseless coding theorem \cite{schumacher} marked a turning point in the field. This shift from focusing on classical information processed by quantum systems to treating quantum information in its own right fundamentally changed how we approach quantum phenomena. The parallel with Shannon's classical information theory \cite{shannon} provides a familiar framework for computer scientists and engineers, that can make quantum concepts more accessible to non-physicists. Equally significant is the shift in emphasis in quantum measurements from physical observables measured to measurement bases. This reframing, as Educator 2 noted, represents a fundamental difference in how quantum information scientists think about measurement compared to traditional quantum physicists. While a physicist might focus on measuring the Z-component of spin with specific eigenvalues, a quantum information scientist cares primarily about the measurement basis itself regardless of the underlying observable that is measured. This abstraction removes unnecessary physical details while preserving the essential information-theoretic content making it easier for QIST educators to be in the ZPD of students from non-physics disciplinary backgrounds. However, as Educator 3 observed, the language of QIST continues to evolve rapidly, with new terms like ``magic in gate” (referring to a gate’s ability to generate entanglement) emerging as researchers develop new concepts. This linguistic explosion, while potentially confusing, reflects the creative growth in a rapidly developing field. The challenge for educators is to navigate this evolving landscape while providing students with a stable conceptual foundation.

\subsection{Navigating Layers of Abstraction vs. Physical Details}

The question of appropriate abstraction levels vs. physical details in QIST education reveals a fundamental tension between accessibility and deep understanding involving physical basis. The responses to the question about entanglement and quantum logic gate operations highlight some quantum educators' views on how students at different levels of the ``stack” in this interdisciplinary field may require different depths of understanding. Also, the answer to this question depends on the scope and goals of the QIST course, and even in different QIST foundations courses, there may be different mathematics and physics prerequisites so a different balance of abstraction and physical details may be appropriate. Educator 3's analogy to classical computing is particularly illuminating. Just as Python programmers need not understand CPU cache architecture, quantum algorithm developers may not need to understand the physical mechanisms underlying quantum gates. This layered abstraction model, familiar to computer scientists, provides a lens for thinking about QIST education. Students working at the application layer can treat quantum computers as black boxes that execute quantum circuits, while those working on hardware development or those involved in error correction in physical systems need deeper physical understanding. However, several educators emphasized the importance of maintaining some connection to physical reality even at higher abstraction levels. Educator 7’s approach of teaching the physical basis of quantum operations, e.g., explaining exchange interactions through wave function overlap and the Pauli exclusion principle, ensures that students understand not just what quantum gates do, but why they work. This grounding in physical principles provides students with the conceptual tools to reason about quantum systems even when working at higher abstraction levels. The challenge, as Educator 4 noted, is that different disciplines bring different comfort levels with abstraction. Computer scientists, trained in thinking about Turing machines and computational complexity, may be comfortable treating quantum computers as abstract information processors. Physicists, who are accustomed to thinking about specific physical systems and their dynamics, may find this level of abstraction unsatisfying or incomplete. A framework for QIST education can lead to a more effective design of QIST courses and bridge these disciplinary differences while maintaining rigor and conceptual clarity to ensure that instruction is within the ZPD of students from diverse interdisciplinary backgrounds.

\subsection{Pedagogical Strategies for Interdisciplinary Education}

The educators interviewed offered several concrete strategies for effective interdisciplinary QIST teaching. Educator 5’s emphasis on finite-dimensional Hilbert spaces over wave mechanics, which is typical for physics majors, represents a fundamental shift in presenting quantum concepts to students from different disciplinary backgrounds to remain in their ZPD. By focusing on linear algebra and matrix representations rather than differential equations, instructors can make quantum concepts accessible to students with diverse mathematical backgrounds while preserving the essential physics. The collaborative teaching model described by Educator 2, in which physics and mathematics faculty jointly taught a course on protecting information, demonstrates how interdisciplinary expertise can be leveraged in the classroom. By having each instructor contribute their disciplinary strengths, e.g., physicists providing physical intuition, mathematicians emphasizing rigorous proofs, students from different disciplinary backgrounds can benefit from multiple perspectives while seeing how different approaches complement each other. Some educators stressed the importance of actively engaging with colleagues from other disciplines to understand their perspectives and challenges to help students learn these interdisciplinary QIST concepts. The cross-disciplinary dialogue is essential for developing an effective framework for QIST education and pedagogical approaches that serve all students well while remaining in their ZPD. The comparison by Educator 6 of interdisciplinary QIST teaching with teaching physics courses for non-science majors such as Physics for Poets is particularly apt. Just as physicists must adjust their language and emphasis when teaching non-majors and it is possible to teach these types of courses effectively if they focus on concepts instead of mathematical formalism, QIST educators must calibrate their level of presentation to their interdisciplinary audience while maintaining conceptual focus to remain in students' ZPD. The challenge, as Educator 6 noted, shifts from being primarily about mathematical formalism to being about presenting concepts in accessible ways.

\subsection{Preparing Students for an Uncertain but Exciting Future}

The unanimous enthusiasm of the educators regarding career prospects in QIST reflects both the current opportunities and the transformative potential of the field. Educator 7’s observation that students have a ``unique position in history” to participate in creating something fundamentally new captures the excitement that draws many to the field. Unlike established disciplines where practitioners primarily work within existing frameworks, QIST offers the opportunity to build entirely new technologies and theoretical frameworks. However, quantum educators also emphasized the importance of developing transferable skills and maintaining adaptability. Educator 4's candid acknowledgment that the ``QIST bubble” may eventually burst, combined with their emphasis on the valuable synthesis of skills students should develop, provides a realistic perspective. The interdisciplinary nature of QIST, which requires understanding physics, mathematics, computer science, chemistry, and engineering, equips students with a broad skill set that will serve them well regardless of how the field evolves from these early days.
The rapid growth of industry opportunities noted, e.g., by Educator 5, demonstrates that QIST as a field is striving to 
realize practical applications of quantum technologies. The fact that students can find positions in both academia and industry, working on problems ranging from fundamental physics to practical engineering challenges, speaks to the rapid advancement of the field.

\subsection{Broader Implications for QIST Education}

Our findings suggest several guidelines for developing a framework for effective interdisciplinary course and curriculum design for QIST education. For example, educators should embrace conceptual unification in QIST course design. The qubit concept, measurement basis, quantum logic gates, circuit diagrams and other evolutions and adaptations from an information-theoretic and applied perspective can provide a common language that can unite students from diverse disciplinary backgrounds. Educators should leverage these unifying concepts while acknowledging their limitations, e.g., lack of physical details. However, since the field is in its infancy but evolving rapidly, the common language will keep evolving to adapt to the needs of stakeholders from diverse disciplinary backgrounds. It would also be valuable to tailor abstraction levels vs. details of the underlying physics based upon student needs even in a QIST foundations course. Different students need different levels of physical detail depending on their goals and backgrounds. Course design should explicitly consider where students sit in the ``stack” and what level of understanding they need for their intended applications. For example, some educators had noted that students in the same course can be given different types of projects to work on collaboratively, and those with a greater interest in physical basis can focus on those issues. Moreover, effective QIST education requires ongoing communication between educators from different disciplines, so strategies for fostering interdisciplinary dialogue would be valuable for developing a framework for QIST education. Joint teaching, cross-disciplinary collaboration, and regular dialogue about pedagogical challenges can improve outcomes for all students. While maintaining mathematical and conceptual rigor, educators should balance the rigor with accessibility and strive for level of presentations that minimize unnecessary complexity. Finite-dimensional systems, pedagogical use of linear algebra as needed, and careful use of visualization can make quantum concepts more accessible to students from diverse backgrounds without sacrificing depth. Furthermore, both educators and students must embrace the rapidly evolving nature of QIST and recognize the basic tenets of QIST, i.e., the ability to adapt, learn new concepts, and learn to work across disciplinary boundaries are critical.

\section{Summary, Conclusions and Future Directions}

The emergence of quantum information science and technology as a truly interdisciplinary field presents both opportunities and challenges for educating students. Through interviews with leading quantum researchers who are actively engaged in teaching, we have identified some themes and strategies that can be valuable for 
developing a framework for QIST education. The development of a common conceptual language, centered on concepts like qubits and measurement bases, can provide a foundation for interdisciplinary communication that will continue to evolve. However, educators must carefully navigate the tension between abstraction and physical basis of understanding QIST concepts, ensuring that students develop appropriate levels of insight for their intended roles in the field. Even in QIST foundations courses, depending upon the course prerequisites (e.g., level of prior physics and math requirements and students' disciplinary backgrounds), it is important to deliberate appropriate balance of abstraction and physical details. Some educators suggested that giving students collaborative course projects in which students have the flexibility to choose projects with different levels of physical details can give students from different disciplinary backgrounds opportunities to delve into these issues at a level commensurate with their interest. Successful interdisciplinary QIST education requires a fundamental shift from traditional quantum mechanics pedagogy for physics majors emphasizing the physical basis as well as infinite dimensional Hilbert space(s). By emphasizing finite-dimensional Hilbert spaces, leveraging interdisciplinary teaching teams, and maintaining ongoing dialogue across disciplinary boundaries, educators can create learning environments that serve students from different disciplinary backgrounds equally well. 
The educators' enthusiasm for the field and their encouragement for students to pursue QIST careers reflect the unique historical moment we inhabit. As we stand at the beginning of the second quantum revolution, students have the opportunity to contribute to transformative quantum technologies while developing valuable interdisciplinary skills that will serve them throughout their careers. 

These findings can provide a foundation for QIST curriculum developers and individual instructors so that students from diverse disciplinary backgrounds are successful in charting their educational paths in this exciting field. By embracing the interdisciplinary nature of this field,
we can prepare the next generation of quantum scientists and engineers to realize the full potential of quantum technologies. The insights shared by these experienced educators provide valuable guidance for educators and researchers involved in the development of a framework to guide course and curriculum development for QIST education at this early time in the second quantum revolution. As QIST continues to evolve, we will need to fine tune our educational approaches to ensure that students from interdisciplinary backgrounds in QIST courses continue to learn these exciting concepts. 
For example, it is possible that after many decades if quantum technologies mature, less physical basis may be appropriate for various concepts in QIST courses at a certain level. However, currently the educators felt that discussion of some physical basis is important since QIST is in its infancy.

The findings presented here can be valuable for developing a framework for QIST education and applying the framework to develop courses and curricula. 
It would be valuable to deliberate how we can effectively assess student understanding across different abstraction levels. It is also valuable to consider the kinds of questions and problems that best reveal different aspects of QIST learning, e.g., students' conceptual understanding and their problem solving, reasoning, and meta-cognitive skills, in the context of QIST courses with a given set of prerequisites. 
It would also be valuable to investigate the optimal sequence for introducing QIST concepts to students from different disciplinary backgrounds and whether all students should start with the same foundational concepts, and whether their pathways should diverge at some point based on the disciplinary background and interests. And if different pathways are appropriate, at what point in the curriculum would it be appropriate for students with different backgrounds and interests within QIST to follow different pathways? It would also be useful to investigate how educators can better understand quantum industry needs and ensure that students are prepared for both current and future workforce demands in QIST. Additionally, it would be useful to investigate how different educational systems and cultural contexts approach QIST education and what we can learn from international comparisons.

\section*{Acknowledgments}
We are grateful to all educators who participated in this research. 
We also thank Educator 2 for allowing us to include the history of how the concept and language of quantum teleportation were developed, which is presented in the Appendix and identifies Educator 2.

\section*{Disclosure statement}

No potential conflict of interest was reported by the authors.

\section*{Authors’ contributions}
L.D., F.S. and C.S. contributed to analysis and interpretation of data, as well as writing and revision of the manuscript. C.S. contributed to the conception and design of research. L.D., F.S. and C.S. contributed to acquisition and interpretation of data, and revision of the manuscript. All authors read and approved the final manuscript.

\section*{Funding}

This research is supported by the US National Science Foundation Award PHY-2309260.

\section*{Availability of data and materials}
The datasets used and analyzed during the current study are not available due to confidentiality agreement with interviewees.

\section*{Ethical Statement}
This research was carried out in accordance with the principles outlined in the University of Pittsburgh 
Institutional Review Board (IRB) ethical policy, the Declaration of Helsinki, and local statutory requirements. The educators provided consent for use of the interview data for research and publication, and consent for quotes to be used.

\section*{Appendix: Historical Development of the Physics of Quantum Teleportation including Coining of the phrase ``Quantum Teleportation"}

Educator 2 described how they, along with their five collaborators, invented quantum teleportation, which is also important from the historical perspective. They said, “It's easy to say how that collaboration came about. Gilles Brassard, who is at the University of Montreal, invited me there in 1992 to give a talk on some work I had done with Asher Peres. What Asher and I had shown is this: suppose you have 2 quantum systems separated by some distance, and suppose they have the same quantum state, but you don't know which state they have and you're trying to figure out what state they have---you're trying to extract classical information from this pair of objects---then we showed that sometimes, under certain circumstances, it's better to make a joint measurement on the pair than to measure them separately, even if you have classical communication. So even measuring one of them and then communicating to the other, then measuring the other and going back and forth, is not as good as making a true joint measurement on the pair.  In the talk at the University of Montreal, I said it's better to bring the two objects together to make a joint measurement, and you can learn more like that than you can by measuring them separately. And at the end of the talk in the question session, Charlie Bennett asked, do you really have to bring them together? If you have lots of entanglement between these 2 places, is there some way you can use that entanglement to make your measurement without actually bringing the 2 objects together?...I didn't know the answer to Charlie's question...So over lunch and for the rest of the day, Charlie and Gilles and I and two others who were in the audience, Claude Crépeau and Richard Jozsa, talked about this problem, thinking about how we could do the joint measurement if we had entanglement. I'm not sure I remember all the details. People made various suggestions, but we didn't have the answer by the end of the day.  Then I had to go back to Williamstown, and I think Charlie had to go back to Yorktown Heights.  So we continued the discussion by email.  We included Asher in the email discussion. He was in Haifa, Israel. And at some point---it didn't take long---we figured out, oh, okay, this is how you do it. You use entanglement to teleport the state of one of the particles to the location of the other one, and then you measure them together. And it turns out that the teleportation is even more interesting than the joint measurement. So, then the 6 of us wrote up a paper…Now, regarding how we all happened to be in Montreal: Gilles had invited Charlie and Claude to visit (Claude was based in Paris). And Richard was employed at the University of Montreal at the time. So that's how the collaboration came about. Five of us were together in Montreal, and it made sense to bring Asher into the discussion since I was talking about work I had done with him…It really does help to have people talking to each other, because someone will suggest an idea, another will say yeah, that's just not gonna work. Well, maybe it doesn't work but that idea will suggest another idea, and so on.” This was the story of how quantum teleportation, a concept and phrase Educator 2 helped coin, was originally envisioned.

\section{References}

\bibliography{refs}
\end{document}